\documentclass[conference]{IEEEtran}
\IEEEoverridecommandlockouts

\usepackage{cite}
\usepackage{amsmath}
\usepackage{amssymb}
\usepackage{algorithm}
\usepackage{algorithmic}
\usepackage{graphicx}
\usepackage{xspace}
\usepackage[T1]{fontenc}
\usepackage{microtype}
\usepackage[htt]{hyphenat}
\usepackage{hyperref}
\usepackage{orcidlink}
\hypersetup{pdftitle={Know Your Contract: eIDAS-Based Verifiable Legal Identities for Smart Contracts}, pdfauthor={Vaziry, Rodriguez Garzon, Wronka, Kuepper}}

\newcommand{\one}{({\textit{i}})\xspace}
\newcommand{\two}{({\textit{ii}})\xspace}
\newcommand{\three}{({\textit{iii}})\xspace}
\newcommand{\four}{({\textit{iv}})\xspace}

\begin{document}

\title{Know Your Contract: eIDAS-Based Verifiable Legal Identities for Smart Contracts, Enabling Regulatory-Compliant On-Chain Operations}

\author{
\IEEEauthorblockN{%
Awid Vaziry\,\orcidlink{0009-0007-2192-5968}\IEEEauthorrefmark{1},
Sandro Rodriguez Garzon\,\orcidlink{0000-0001-6921-294X}\IEEEauthorrefmark{1},
Christoph Wronka\,\orcidlink{0000-0001-5074-433X}\IEEEauthorrefmark{2},
Axel Küpper\,\orcidlink{0000-0002-4356-5613}\IEEEauthorrefmark{1}}
\IEEEauthorblockA{\IEEEauthorrefmark{1}Service-centric Networking / T-Labs, Technische Universität Berlin, Germany\\
\{vaziry, rodriguezgarzon, axel.kuepper\}@tu-berlin.de}
\IEEEauthorblockA{\IEEEauthorrefmark{2}Financial Services Audit \& Advisory, Baker Tilly, Hamburg, Germany\\
christoph.wronka@bakertilly.de}
}

\maketitle
\bstctlcite{IEEEexample:BSTcontrol}

\begin{abstract}
Public blockchains provide no native mechanism to verify the legal identity behind a deployed smart contract, which blocks institutional adoption and compliance with EU regulations such as MiCA and AMLR. We present \emph{KYC~Seal}, the first protocol that extends the EU eIDAS trust infrastructure to Ethereum smart contracts by cryptographically binding them to Qualified Electronic Seals issued by Qualified Trust Service Providers (QTSPs). The protocol realizes the full eIDAS trust chain, from the European Commission's List of Trusted Lists through Member-State trusted lists and QTSP-signed X.509 certificates down to the individual smart contract, natively on-chain. An on-chain parser extracts identity fields directly from the QTSP-signed certificate bytes at registration. Both cryptographic verifications, the QTSP issuance signature and the certificate holder's seal signature, are performed once at registration and cached as on-chain state, reducing per-interaction seal verification to a pure state check. A new P-256 elliptic-curve precompile in Ethereum (deployed December~2025) makes these one-time cryptographic steps economical, enabling trustless on-chain verification of eIDAS identities without oracles or runtime intermediaries. A reference implementation, a formal security analysis, and a gas evaluation are the subject of forthcoming work.
\end{abstract}


\section{Introduction}
\label{sec:introduction}

Public blockchain ecosystems execute transactions between counterparties whose legal identities cannot be derived from on-chain data: account addresses are pseudonymous identifiers without verifiable linkage to real-world persons or organisations. While pseudonymity supports privacy and resists censorship, it creates substantial challenges for regulated applications. Traditional Finance (TradFi) institutions and applications involving tokenised Real-World Assets (RWAs) require transactions to be attributable to legally accountable entities~\cite{oecd2020tokenisation}. Without such attribution, regulatory compliance and legal certainty cannot be established.

A central obstacle to institutional adoption of public blockchains is the lack of a verifiable trust anchor linking on-chain accounts and smart contracts to real-world legal entities~\cite{vaziry_sok_2024}. Financial regulation requires transactions to be attributable, auditable, and traceable to responsible persons or organisations, yet blockchain systems provide no native mechanism for such attribution~\cite{wronka2024crypto}. Existing digital identity infrastructures offer no standardised way to project certified organisational identities into decentralised environments. Current on-chain identity approaches rely on off-chain verification, proprietary intermediaries, or unverifiable attestations, resulting in fragmented trust models, increased compliance risk, and single points of failure~\cite{vaziry_sok_2024}. Consequently, enterprises lack an interoperable and regulator-recognised method to assert their legal identity within public blockchain ecosystems.

The Ethereum Fusaka upgrade (December~2025) introduced a P-256 precompile (EIP-7951~\cite{eip_7951_2025}) that makes cryptography on the eIDAS-mandated curve economical on the EVM for the first time. Building on this tipping point, we present \emph{KYC~Seal}, a protocol that extends the EU eIDAS trust infrastructure natively into the Ethereum Virtual Machine (EVM). KYC~Seal binds smart contracts to Qualified Electronic Seals (QSeal) issued by Qualified Trust Service Providers (QTSPs), realizing the full trust chain from the European Commission's List of Trusted Lists down to the target contract's address without runtime reliance on oracles or off-chain intermediaries. An on-chain ASN.1/DER parser extracts identity fields directly from QTSP-signed X.509 certificates, closing the semantic gap between signature verification and identity attribution. Both cryptographic verifications are performed once at registration and cached as on-chain state, reducing per-interaction verification to a pure state check suitable for frequent use by compliance-sensitive contracts.

\noindent\textbf{Our Contribution.} We make the following contributions:
\begin{itemize}
\item We design KYC~Seal, the first protocol that extends the EU eIDAS trust infrastructure to EVM smart contracts, enabling trustless on-chain verification of legally recognised identities without oracles or runtime intermediaries.
\item We develop the first \emph{on-chain ASN.1/DER parser} for X.509 certificates that extracts public keys, Legal Person Identifiers, and validity periods directly on the EVM, closing the semantic trust gap between QTSP-signed certificates and on-chain identity binding.
\item We show that performing both cryptographic verifications once at registration and caching them as on-chain state reduces per-interaction seal verification to a pure state check with no per-call signature operation.
\end{itemize}
A reference implementation, a formal security analysis, and an empirical gas evaluation of KYC~Seal are the subject of forthcoming work.

\noindent\textbf{Paper Organization.}
Section~\ref{sec:related-work-pki} surveys related work on blockchain-based PKI and on-chain organisational identity. Section~\ref{sec:background} introduces the eIDAS trust infrastructure, X.509 certificate structure, and the regulatory drivers for on-chain identity. Section~\ref{sec:system-threat-model} defines the system and trust model. Section~\ref{sec:protocol} presents the KYC~Seal protocol. Section~\ref{sec:discussion} discusses limitations and future work, and Section~\ref{sec:conclusion} concludes.

\section{Related Work}
\label{sec:related-work-pki}

\noindent\textbf{Blockchain-Based Public Key Infrastructure.} Blockchain-based PKI research has followed four threads: smart contracts that automatically impose financial penalties on Certificate Authorities (CAs) that issue unauthorised certificates (IKP~\cite{matsumoto_ikp_2017}), transparency logs recording TLS certificates and revocation on-chain (CertLedger~\cite{kubilay_certledger_2019}, Wang et al.~\cite{wang_blockchain_ct_2022}), blockchain-native PKI systems that reimplement issuance, validation, and revocation as smart contracts with CAs operating on-chain (SCPKI by Al-Bassam~\cite{albassam_scpki_2017}, Patsonakis et al.~\cite{patsonakis_smart_contract_pki_2020}, ProofChain by Saleem et al.~\cite{saleem_proofchain_2022}, Yakubov et al.~\cite{yakubov_blockchain_pki_2018}), and zero-knowledge approaches that prove possession of an existing X.509 certificate without revealing its contents (zk-X509~\cite{bak_zkx509_2026}). The first three approaches are \emph{CA-rooted} and extract no identity fields from certificate structures within contract execution~\cite{khan_survey_x509_pki_2023}; zk-X509 consumes existing PKI but targets natural-person identity with full cryptographic privacy, hiding certificate contents behind an off-chain proof.
KYC~Seal differs by operating within the eIDAS trust chain, a \emph{PKI-rooted} model whose trust anchor is a multi-authority, government-curated trust-list infrastructure rather than a single CA. It lets Solidity smart contracts verify and consume a legal identity at runtime, which the CA-rooted and zero-knowledge approaches above cannot, establishing legal-person accountability for EU regulatory compliance.

\noindent\textbf{On-Chain Organisational Identity and Trust Frameworks.} GLEIF's verifiable Legal Entity Identifier (vLEI)~\cite{GLEIF_vLEI_repo} encodes legal-person identity as a wallet-held credential governed by GLEIF rather than as an eIDAS trust service.
Bartolomeo~\cite{bartolomeo_functional_2025} developed Functional Credentials, a privacy-preserving construction targeting the EU Digital Identity Wallet for natural persons.
Both are off-chain credential formats whose verification occurs between a wallet and a relying party rather than inside a smart contract.

Chainlink's Automated Compliance Engine (ACE)\footnote{\url{https://blog.chain.link/automated-compliance-engine-technical-overview/}} places identity records on public chains, but the underlying vLEI is parsed and verified off-chain by the Chainlink oracle network and only the resulting attestation is written on-chain.
The European Blockchain Services Infrastructure (EBSI)~\cite{samuel_gomez_european_2025} operates a permissioned ledger whose trust registries (Trusted Entity Registry, Trusted Accreditation Registry) publish governance-backed issuer and accreditation records, while end-user verifiable credentials are held off-chain in holder wallets under the W3C self-sovereign model.
Both systems move compliance data on-chain, but neither verifies a qualified cryptographic artefact inside contract execution on a public chain: ACE trusts its oracle network, and EBSI confines verification to its own permissioned ledger.

The amended eIDAS regulation introduced the qualified electronic ledger as a new trust-service class (Art.~45k, 45l); Alamillo~et~al.~\cite{alamillo_qualified_2024} analyze how such ledgers can provide legal certainty for on-chain records, but the construction presupposes QTSP control of the ledger, positioning it toward permissioned rather than public chains.

\begin{table*}[t]
\centering
\caption{Positioning of KYC~Seal against blockchain-based PKI and on-chain identity approaches.
\textbf{On-chain identity} (the extent to which certificate fields are verified on-chain):
None (no Blockchain used),
Off-chain (verification off-chain, result on-chain),
Permissioned (Yes, but on a restricted ledger),
Selective (zkp / selective disclosure),
Yes (full public-chain verification).
\textbf{Composable} (whether another contract can atomically use the trusted identity data during execution):
Restricted (permissions apply),
Attestation (compliance flag only, no identity data),
Yes (fully queryable at runtime).}
\label{tab:related-work-comparison}
\setlength{\tabcolsep}{4pt}
\resizebox{\textwidth}{!}{%
\begin{tabular}{lccccc}
\hline
\textbf{Approach} & \textbf{Trust model} & \textbf{Intermediary} & \textbf{Regulation} & \textbf{On-chain identity} & \textbf{Composable} \\
\hline
IKP~\cite{matsumoto_ikp_2017}                     & CA-oversight        & --                 & --                 & Off-chain    & No \\
CertLedger~\cite{kubilay_certledger_2019}         & CA-rooted           & --                 & --                 & Off-chain    & No \\
SCPKI~\cite{albassam_scpki_2017}                  & Web-of-trust        & --                 & --                 & Off-chain    & No \\
zk-X509~\cite{bak_zkx509_2026}                    & CA-rooted (ZK)      & --                 & --                 & Selective    & Hashes only \\
vLEI~\cite{GLEIF_vLEI_repo}                       & GLEIF-rooted        & Wallet             & ISO~17442          & None         & No \\
Funct. Cred.~\cite{bartolomeo_functional_2025}    & Wallet-based        & --                 & eIDAS          & None         & No \\
Chainlink ACE                                     & Oracle              & Chainlink          & --                 & Off-chain    & Attestation \\
EBSI~\cite{samuel_gomez_european_2025}            & Governmental        & MS validators      & eIDAS (impl.)  & Permissioned & Restricted \\
Qual.\ ledger~\cite{alamillo_qualified_2024}      & QTSP-controlled     & QTSP               & eIDAS~45k/45l      & Permissioned & Restricted \\
\textbf{KYC~Seal}                     & \textbf{PKI-rooted} & \textbf{--}        & \textbf{eIDAS}     & \textbf{Yes} & \textbf{Yes} \\
\hline
\end{tabular}%
}
\end{table*}

\noindent\textbf{KYC~Seal} (Table~\ref{tab:related-work-comparison}) operates on a genuine eIDAS-qualified artefact (a QTSP-issued X.509 certificate chaining to the EU LOTL), executes full trust-chain verification natively inside an EVM smart contract on a public chain, and binds an individual contract address to a specific legal person. The verified trust chain becomes on-chain state and can be queried within atomic transactions, enabling regulatory-compliant, identified interactions.

\section{Background}
\label{sec:background}
This section introduces the technical and regulatory foundations underlying the KYC Seal protocol: the EU eIDAS trust infrastructure, X.509 certificate structure, and the regulatory drivers for on-chain identity.

\noindent\textbf{Terminology.} Throughout this work, we distinguish between \emph{verification} (the cryptographic operation of checking a signature against a public key) and \emph{validation} (the broader process of confirming a certificate's trust chain, revocation status, and policy compliance).

\subsection{EU eIDAS Trust Infrastructure}
\label{sec:eidas-trust}
The eIDAS Regulation (EU No 910/2014)~\cite{eIDAS1}, operational since 2016 and amended in 2024~\cite{eIDAS2}, provides the EU's trust framework for electronic identification and trust services. Central to this framework are QTSPs, authorised to issue legally recognised trust services including qualified electronic signatures, seals, and attestations of attributes~\cite{eIDAS1}. Member States publish machine-processable Trusted Lists (TSLs) of QTSPs, aggregated by the European Comission (EC) into the List of Trusted Lists (LOTL)~\cite{eIDAS1}. This infrastructure guarantees that certificates can be traced back to a root controlled by the EC, with revocation and status information available in machine-processable form. A qualified electronic seal is the eIDAS term for a digital signature issued by a legal person. For a seal to qualify, it must be created using cryptographic mechanisms that comply with the EC's implementing regulations. The regulatory reference chain begins with implementing regulation 2024/2979~\cite{the_european_parliament_and_the_council_of_the_european_union_commission_2024}, which specifies ETSI standards for seal formats~\cite{ETSI_TS-119-312_2024}. ETSI TS~119~312 defines permitted cryptographic suites including the Elliptic Curve Digital Signature Algorithm (ECDSA) with P-256 (secp256r1) as a recommended algorithm.

\subsection{X.509 Certificate Structure}
\label{sec:x509-structure}
QTSP-issued qualified certificates for electronic seals follow the X.509v3 standard~\cite{rfc5280}, encoded in ASN.1 Distinguished Encoding Rules (DER). Each certificate contains a To-Be-Signed (TBS) Certificate structure holding the subject's public key, Legal Person Identifier (LPID)~\cite{ETSI-EN-319-412-1}, validity period, and issuer information. The QTSP signs this TBS structure to produce the complete certificate. For on-chain verification, two operations are relevant: \one parsing the TBS certificate to extract the public key and identity fields, and \two verifying the QTSP's signature over the TBS certificate against the QTSP's own certificate from the TSL.

\subsection{Regulatory Drivers for On-Chain Identity}
\label{sec:regulatory-drivers}
EU financial regulation requires that value-bearing digital actions be attributable to verifiable legal entities. The Payment Services Directive (PSD2)~\cite{eu_psd2_2015} and its successor, the Payment Services Regulation (PSR)~\cite{european_commission_psr_2023}, mandate unambiguous attribution of payment instructions to legally responsible persons. The Anti-Money Laundering Regulation (AMLR)~\cite{eu_amlr_2024} requires traceability of both initiating and intermediary parties. The Markets in Crypto-Assets Regulation (MiCA)~\cite{eu_mica_2023} requires crypto-asset service providers to identify customers, ensure transaction traceability, and maintain auditability. Together, these regulations create an \emph{attribution gap} on public blockchains: pseudonymous on-chain components can perform regulated actions, but there is no native way to verify which legal entity controls them~\cite{wronka2024crypto}.

\section{System and Trust Model}
\label{sec:system-threat-model}

\begin{figure*}[t]
\centering
\includegraphics[width=0.9\textwidth]{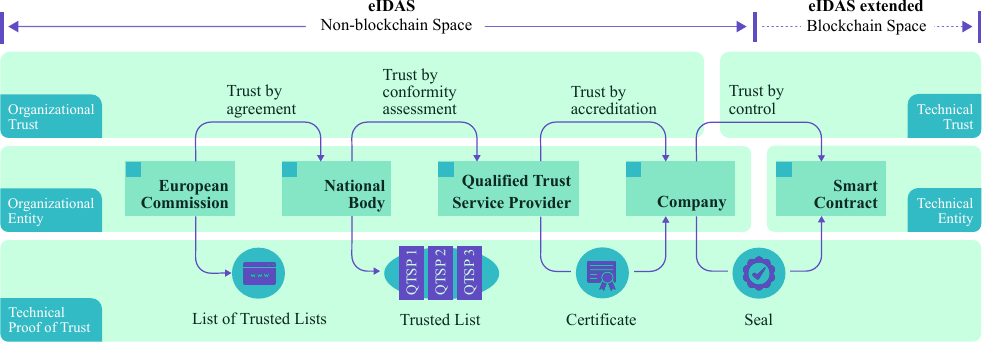}
\caption{Layered model of the EU eIDAS trust chain, showing which entities are involved, why those entities are trusted, and how the trust is realized by technical means. The trust chain extension to the blockchain is then illustrated.}
\label{fig:trust-chain}
\end{figure*}

Figure~\ref{fig:trust-chain} shows the architecture of KYC Seal, which extends the EU eIDAS trust chain onto the blockchain and enables validated on-chain operation without external oracles at runtime. The main trust assumption is the mirroring of the LOTL into the on-chain \texttt{TrustedList}, both of which are governance-curated. All subsequent artifacts and operations are verified and validated against this mirrored trust chain, with \texttt{TrustedList} as the root of trust. During the registration of a new qualified certificate or qualified seal, it becomes trusted once verified against the TL during transaction execution and is then persisted in its trusted form on-chain until revocation.

\subsection{System Model}
\label{sec:system-model}

The system model comprises three main smart contracts, three cryptographic artifacts, and five actors, connected by the trust chain of Figure~\ref{fig:trust-chain}. The remainder of this section describes each element.

\noindent\textbf{Contracts and Artifacts.} The main contracts are \texttt{TrustedList}, which serves as the root trust store for qualified trust services; \texttt{SealCertificateRegistry}, which stores and manages the sealing certificate and its issued seals; and \texttt{SealedContract}, a base contract designed to be inherited by any contract that requires sealing. The three cryptographic artifacts are the QTSP's service digital identity (\texttt{SDI}, clause~5.5.3 of ETSI~TS~119~612), the issuer-side certificate rooted in the LOTL that issues end-entity seal certificates; the qualified certificate for electronic seal (\texttt{QSealC}, clause~3.3 of ETSI~TS~119~495), the end-entity certificate bound to the legal person per eIDAS~Art.~3(29)--(30); and the QSeal itself, represented by its TBS data and additional fields extracted from the CAdES envelope and committed raw on-chain.

\noindent\textbf{Actors.}
Five actors participate: the \emph{Trust Root Governance} curates the on-chain trust store; \emph{QTSPs} sign certificates off-chain under eIDAS vetting; the \emph{Contract Owner} approves a specific \texttt{certId} for sealing via \texttt{onlyOwner}; the \emph{Certificate Holder} signs the seal payload off-chain with the \texttt{QSealC} private key; and the \emph{Verifier} calls \texttt{verifySeal} without any trust requirement. Seal registration is authorised cryptographically by the Certificate Holder's P-256 signature over the seal payload rather than by \texttt{msg.sender}; transaction submitters are therefore not a distinct trust principal.

\section{KYC Seal Protocol}
\label{sec:protocol}
\begin{algorithm}[t]
\caption{\texttt{registerCertificate}($\mathsf{fullCertDER}$, $\mathsf{tsId}$)}
\label{alg:register-cert}
\begin{algorithmic}[1]
\REQUIRE DER-encoded \texttt{QSealC} $\mathsf{fullCertDER}$; trust service identifier $\mathsf{tsId}$
\ENSURE Certificate record stored; $\mathsf{certId}$ returned
\item[] \textit{Step~1: Preconditions and identifier derivation}
\STATE \quad \textbf{require} $\textsc{IsValidTrustService}(\mathsf{tsId})$
\STATE \quad $\mathsf{rawTBS} \gets \textsc{ExtractTBS}(\mathsf{fullCertDER})$
\STATE \quad $\mathsf{certId} \gets \texttt{sha256}(\mathsf{rawTBS})$
\STATE \quad \textbf{require} $\neg\,\textsc{Exists}(\mathsf{certId})$
\item[] \textit{Step~2: Verify QTSP issuer signature (V2)}
\STATE \quad $(r_{\mathrm{cert}}, s_{\mathrm{cert}}) \gets \textsc{ExtractSignatureRS}(\mathsf{fullCertDER})$
\STATE \quad $s_{\mathrm{cert}} \gets \textsc{NormalizeLowS}(s_{\mathrm{cert}})$
\STATE \quad $(\mathsf{ts}_x, \mathsf{ts}_y) \gets \textsc{GetTrustServiceKey}(\mathsf{tsId})$
\STATE \quad \textbf{require} $\texttt{P256.verify}(\mathsf{certId}, r_{\mathrm{cert}}, s_{\mathrm{cert}}, \mathsf{ts}_x, \mathsf{ts}_y)$
\item[] \textit{Step~3: Parse TBS and persist record}
\STATE \quad $(\mathsf{pk}_x, \mathsf{pk}_y, \mathsf{lpid}, \mathsf{t}_{\min}, \mathsf{t}_{\max}, \mathit{serial}) \gets \textsc{ParseCertTBS}(\mathsf{rawTBS})$
\STATE \quad \textbf{require} $\mathsf{t}_{\min} \leq \texttt{block.timestamp} \leq \mathsf{t}_{\max}$
\STATE \quad Store $(\mathsf{pk}_x, \mathsf{pk}_y, \mathsf{lpid}, \mathsf{t}_{\min}, \mathsf{t}_{\max}, \mathsf{tsId}, \mathit{serial}, \mathsf{true})$ under $\mathsf{certId}$
\STATE \quad \textbf{emit} $\textsc{CertificateRegistered}(\mathsf{certId}, \mathsf{lpid}, \mathsf{tsId})$
\RETURN $\mathsf{certId}$
\end{algorithmic}
\end{algorithm}
This section specifies the \emph{KYC Seal} protocol, which extends the eIDAS trust onto the blockchain, as discussed in Section~\ref{sec:system-model}. The protocol decomposes into two phases. \textbf{Phase~1 (Registration)} establishes the trust chain in on-chain state through five one-time steps: (1.1)~trust anchor provisioning, (1.2)~certificate registration, (1.3)~seal approval, (1.4)~off-chain sealing, and (1.5)~on-chain seal registration. Their cost is incurred once and then reused: trust anchor provisioning~(1.1) and certificate registration~(1.2) amortise over all downstream seals, whereas seal approval, sealing, and registration~(1.3--1.5) run once per sealed contract. \textbf{Phase~2 (Usage)} is the recurring read path: a single step, (2.1)~seal verification, re-evaluated on every guarded interaction. Because both P-256 signature verifications (V1, the certificate holder's seal signature; V2, the QTSP issuance signature) are performed during Phase~1 and cached as on-chain state, Phase~2 reduces to a pure state check with no per-interaction P-256 operation.
The protocol binds a legal identity, vetted by EU regulation, to an on-chain smart contract at address $\mathsf{target}$ with the chain ID $\mathsf{cid}$.

\subsection{Trust Anchor and Certificate Registration}
\label{sec:registration}
The trust anchor \texttt{SDI} certificate has a 1 to n relationship to \texttt{QSealC}, which itself has a 1 to n relationship to issued Seals. The computational and storage costs for the following registration steps can therefore be amortised over various child certificates and Seals respectively.

\smallskip
\noindent\textbf{Step~1.1 (Trust Anchor Provisioning).} Making the trusted QTSP certificate of the LOTL available on-chain in the \texttt{TrustedList}, which acts as the on-chain root trust store. The Trust Root Governance actor (for example, an EU institution or an Ethereum Foundation governance body) mirrors the full DER-encoded \texttt{SDI} of each relevant QTSP from the web-published LOTL into \texttt{TrustedList} via \texttt{registerTrustService}. Thereby the TBS is parsed to extract the public key of the QTSP, validity window, and organisational identifier, indexed by $\mathsf{tsId} = \mathtt{sha256}(\mathsf{rawTBS})$ to prevent re-registration. No issuer signature is verified, since trust is governance-curated, just as browser and operating-system root stores ship pre-vetted CA certificates that applications trust implicitly.

\noindent\textbf{Step~1.2 (Certificate Registration).} Any party may submit an end-entity \texttt{QSealC} certificate for registration. The registry contract first extracts the \texttt{TBSCertificate} and derives its identifier $\mathsf{certId}=\texttt{sha256}(\mathsf{rawTBS})$. It then verifies the QTSP issuance signature (V2) using the public key of the issuing \texttt{SDI} already stored in \texttt{TrustedList}. Only if this signature is valid does it parse the identity fields from the signed TBS and store the certificate record on-chain, as detailed in Algorithm~\ref{alg:register-cert}.

The QTSP issuance signature on a qualified seal certificate is ECDSA over $\mathtt{sha256}(\mathsf{TBSCertificate})$~\cite{ETSI_TS-119-312_2024}. Hence the \texttt{P256.verify} \textbf{require} in Step~2 verifies $(r_{\mathrm{cert}}, s_{\mathrm{cert}})$ against $\mathsf{certId}=\mathtt{sha256}(\mathsf{rawTBS})$, which is equivalent to verifying the signature over the submitted TBS bytes. This matches the X.509/ETSI signature semantics and introduces no additional hashing step beyond the standard certificate digest.

By direct inspection of Algorithm~\ref{alg:register-cert}, a returned $\mathsf{certId}$ implies that $\mathsf{tsId}$ is active and that the QTSP issuance signature verifies for $\mathsf{certId}=\mathtt{sha256}(\mathsf{rawTBS})$ under the \texttt{SDI} public key (the two \texttt{require} statements). The public key, LPID, validity window, and serial are then parsed from the \emph{same} $\mathsf{rawTBS}$ bytes, so no identity field is accepted as independent caller input; under correct parsing, the stored record is bound to the very \texttt{TBSCertificate} authenticated by the QTSP signature.

\subsection{Seal Approval, Off-Chain Sealing, and Registration}
\label{sec:seal-lifecycle}

\noindent\textbf{Step~1.3 (Seal Approval).} Before a Seal is registered, the target contract, which inherited \texttt{SealedContract}, must pre-approve sealing by giving consent to seal registration by a specific \texttt{QSealC}. The Contract Owner of a \texttt{SealedContract} calls \texttt{approveCertificateForSealing}($\mathsf{certId}$), which validates the full certificate trust chain at approval time and records $\texttt{sealApprovals}[\mathsf{target}] = \mathsf{certId}$ in \texttt{SealCertificateRegistry}.

\noindent\textbf{Step~1.4 (Off-Chain Sealing).}  The Certificate Holder assembles the CAdES envelope~\cite{ETSI-EN-319-122} carrying the sealed data, computes $\mathsf{envelopeHash} = \texttt{keccak256}(\mathsf{cadesBlob})$, and signs the seal payload with the \texttt{QSealC} private key off-chain, producing $(r_s, s_s)$. The domain separator provides protocol versioning analogous to EIP-712. With $\mathsf{domSep} = \mathtt{keccak256}(\text{``KYCSeal-v1''})$, the seal payload is:
\begin{equation}
\label{eq:seal-payload}
\begin{aligned}
\mathsf{payload} = \mathtt{keccak256}\bigl(&\mathtt{abi.encode}(\mathsf{domSep}, \mathsf{target}, \mathsf{cid},\\
&\mathsf{cert.lpid}, \mathsf{envelopeHash})\bigr).
\end{aligned}
\end{equation}
Binding $\mathsf{envelopeHash}$ into the V1 payload cryptographically commits the holder to a specific CAdES envelope: any later registration that substitutes a different envelope fails V1 verification, reducing envelope-substitution to an ECDSA forgery against $\mathsf{cert.pk}$. Rotating the envelope (for example, a BES $\to$ LT upgrade) therefore requires a fresh signature and a fresh \texttt{registerSeal} call.

\noindent\textbf{Step~1.5 (On-Chain Seal Registration).} Any Ethereum account may call \texttt{registerSeal}($\mathsf{target}$, $\mathsf{sealData}$) with $\mathsf{sealData} = (r_s, s_s, \mathsf{certId}, \mathsf{envelopeHash})$. The registry re-checks the approval, re-validates the full trust chain, verifies V1 against the registered $(\mathsf{cert.pk}_x, \mathsf{cert.pk}_y)$, and atomically stores the seal record $(r_s, s_s, \mathsf{certId}, \mathsf{chainIdAtIssuance}, \mathsf{envelopeHash})$ together with $\mathsf{active} = \mathsf{true}$. Authorisation is entirely cryptographic (V1 P-256 signature), not sender-based, enabling account abstraction and relayer submission.

\subsection{Seal Verification}

\noindent\textbf{Step~2.1 (Seal Verification).} Seal usage is read-only: V2 is verified once at certificate registration and V1 once at seal registration. Subsequent use therefore reduces to loading the stored seal and certificate records and re-evaluating dynamic validity conditions. Algorithm~\ref{alg:verify-seal} specifies this check. The \texttt{onlySealed}($\mathsf{addr}$) modifier exposed by \texttt{SealedContract} delegates to \texttt{SealCertificateRegistry.verifySeal}.

\begin{algorithm}[tbh]
\caption{\texttt{verifySeal}($\mathsf{target}$) / \texttt{onlySealed}($\mathsf{counterparty}$)}
\label{alg:verify-seal}
\begin{algorithmic}
\REQUIRE Target contract address $\mathsf{target}$ with a registered seal record
\ENSURE Returns $\mathsf{true}$ if and only if the registered seal is still active, its certificate's trust chain is intact, and its chain binding matches
\item[] \textbf{Step~1: Load seal and verify chain binding}
\STATE \quad $\mathsf{seal} \gets \texttt{seals}[\mathsf{target}]$
\STATE \quad \textbf{require} $\mathsf{seal.active} \;\wedge\; \mathsf{seal.chainIdAtIssuance} = \texttt{block.chainid}$
\item[] \textbf{Step~2: Load certificate and verify lifecycle}
\STATE \quad $\mathsf{cert} \gets \texttt{certificates}[\mathsf{seal.certId}]$
\STATE \quad \textbf{require} $\mathsf{cert.active} \;\wedge\; \mathsf{cert.pk}_x \neq 0$
\STATE \quad \textbf{require} $\mathsf{cert.t}_{\min} \leq \texttt{block.timestamp} \leq \mathsf{cert.t}_{\max}$
\item[] \textbf{Step~3: Re-check trust anchor}
\STATE \quad \textbf{require} $\textsc{IsValidTrustService}(\mathsf{cert.tsId})$
\RETURN $\mathsf{true}$
\end{algorithmic}
\end{algorithm}
No per-call P-256 verification is needed. A call succeeds if the stored seal is active, was issued for the current chain, and references a certificate that is still active, time-valid, and anchored in a currently valid trust service. Revocation may invalidate either the seal or the certificate. Hence verification is a pure state check.

\subsection{On-Chain ASN.1/DER Parser}
\label{sec:der-parser}

The on-chain parser closes the gap between ``the QTSP signed a hash'' and ``the signed bytes contain the claimed identity fields.'' During certificate registration, the contract extracts the public key, LPID, validity window, and serial directly from the submitted \texttt{TBSCertificate}, i.e., from the same byte string whose SHA-256 digest is verified under the QTSP issuer signature.

The parser performs a single linear walk over the \texttt{TBSCertificate} and extracts:
\begin{itemize}
\item the P-256 public key from \texttt{SubjectPublicKeyInfo};
\item the LPID from \texttt{organizationIdentifier} (OID~2.5.4.97), validated against the \texttt{LEIXG-} prefix;
\item the validity window; and
\item the certificate serial for CRL-based revocation.
\end{itemize}

The parser is intentionally narrow: it accepts only the eIDAS QSeal profile used in our protocol and reverts on malformed or unexpected encodings. We do not claim a formal proof of parser correctness and treat the parser as part of the trusted computing base.

\subsection{Revocation}
\label{sec:revocation}
We define three revocation paths to clear or deactivate a seal: \one~the Contract Owner removes its own seal via \texttt{removeSeal}, \two~the Certificate Holder signs a revocation payload and submits it via \texttt{revokeSealByIssuer}, which verifies the P-256 signature and clears the seal, and \three~any party submits a DER-encoded CRL signed by the \texttt{SDI}, whose serial listing \texttt{CRLParser} searches to permanently set $\mathsf{cert.active} = \mathsf{false}$. OCSP and CRL Distribution Points are universally deployed by EU QTSPs, so coverage is preserved. The issuer revocation payload binds $(\mathsf{target}, \mathsf{cid}, \mathsf{certId})$ without a nonce. Post-revocation replay is defeated by the standard eIDAS lifecycle of certificate rotation.

\section{Discussion}
\label{sec:discussion}

\noindent\textbf{Registration-once, verify-always.}
The steady-state verification path performs \emph{zero} P-256 operations: both signatures are verified exactly once, V2 at \texttt{registerCertificate} and V1 at \texttt{registerSeal}, and cached as unforgeable on-chain state. Monotonic deactivation of the stored seal flag guarantees it cannot be re-established without a fresh V1 check, so expensive cryptographic work is incurred once rather than by every downstream verifier.

\smallskip
\noindent\textbf{Deferred verification as a reusable pattern.}
The two-layer caching used here is not specific to eIDAS. Any protocol whose trust artifact is signed by a slowly changing authority and consumed repeatedly by many parties can follow the same pattern: verify once at registration, persist the result as monotone state, and reduce the steady-state path to bounded storage reads. The pattern shifts verification cost from the verifier to the registrant and decouples the hot path from signature-scheme cost, which is particularly valuable on platforms where new curves arrive only through protocol upgrades.

\smallskip
\noindent\textbf{Generalisation beyond QSeal.} Under eIDAS, Qualified Certificates for Electronic Signatures (QES) and for Website Authentication (QWAC) share the same profile family and the same QTSP-to-LOTL trust chain as QSealC. The identity fields (public key, LPID for QSeal and QWAC, natural-person identifier for QES) occupy the same positions within the TBS, so the on-chain DER parser extends to QES and QWAC without protocol-level changes. KYC~Seal is therefore a \emph{template for all qualified trust services on the EVM}. The principal obstacle for QES is regulatory rather than cryptographic.

\smallskip
\noindent\textbf{Governance of the on-chain trust root.} Mirroring the LOTL onto the chain is the last remaining trust assumption and the main attack surface. Four governance designs trade off differently. \one~\emph{Admin-curated mirror} (the current prototype), analogous to browser CA stores: simple and auditable, but (semi) centralised. \two~\emph{Protocol-level trust anchor}, with the LOTL embedded in the Ethereum client in the same way browsers ship CA stores: strongest integrity, but tightly coupled to client releases. \three~\emph{Individual trusted list}, where on-chain actors manually select QTSPs and pre-approve them as root trust sources: practical, with the added overhead of individually curating lists and comparing them to counterparties. \four~\emph{EU-operated on-chain publication}, in which the European Commission publishes the authoritative trust-service set directly on-chain: the on-chain list becomes the primary source and the web LOTL a legacy mirror, potentially raising trust \emph{above} the current web-served distribution by substituting consensus finality for HTTPS-transport assumptions. Regardless of the chosen option, the governance surface should be hardened with transparency around root updates, append-only audit logs of trust-list mutations, delayed activation of new trust anchors, and multi-party governance with threshold approval for additions and removals.

\smallskip
\noindent\textbf{Limitations.}
\one~The DER parser is narrow by construction and not formally verified, and is therefore treated as part of the trusted computing base. \two~The curated trust-root assumption reduces on-chain verification to ``governance of the on-chain LOTL mirror is honest,'' paralleling rather than eliminating the trust placed in browser CA stores. \three~The seal binds a single CAdES envelope; legitimate envelope rotation (for example, BES~$\to$~LT) requires a fresh seal signature and a new \texttt{registerSeal} call.

\smallskip
\noindent\textbf{Future work.}
\one~A formally verified DER parser, produced through a proof-carrying ASN.1 toolchain, would remove the parser from the trusted computing base. \two~Seal portability across chains would generalise the current per-chain-ID binding into a signed portability statement, enabling a single legal entity to maintain consistent identity across L1 and L2 without independent re-sealing on each chain. \three~Natural-person extensions (QES) with selective disclosure or zero-knowledge proofs would bring GDPR-sensitive attribution within reach of the same architecture. \four~Standardisation via EU initiatives or ERC proposals.

\section{Conclusion}
\label{sec:conclusion}
We presented \emph{KYC~Seal}, the first protocol that verifies the full EU eIDAS trust chain (from the European Commission's LOTL through Member-State TSLs and QTSP certificates down to an individual smart contract) natively inside EVM execution, closing the attribution gap that has blocked compliance with PSD2/PSR, AMLR, and MiCA on public blockchains. Per-interaction seal verification reduces to a pure on-chain state check, while the one-time registration steps perform the QTSP issuer-signature verification and the on-chain DER parse of the TBS, made economical by the recent P-256 precompile on the EVM.
The on-chain ASN.1/DER parser, the registration-time caching of both verifications, and the trust-list governance model are protocol-level components that generalise beyond QSeal to the wider family of eIDAS qualified trust services. A reference implementation, a formal security analysis, and a gas evaluation are the subject of forthcoming work. We invite ERC-level standardisation of the on-chain trust-list semantics, certificate-record layout, and seal-verification interface exposed by this work.

\bibliographystyle{IEEEtran}
\bibliography{references/references,references/references_AI,references/references_regulation,references/references_P5}

\end{document}